\documentclass[manuscript]{acmart}

\usepackage{booktabs}
\usepackage{enumitem}

\newcounter{designCounter}

\AtBeginDocument{%
  }

\begin{document}

% \title[Charting Gaps Between Clinical Research Needs and LLM Data Abstraction Systems]{Charting Gaps Between Clinical Research Needs and LLM Data Abstraction Systems: A Longitudinal Design Study in Oncology}
% \title[Charting Gaps Between Clinical Research Needs and LLM Data Abstraction Systems]{``I Know Where to Look'': Charting Gaps Between Clinical Research Needs and LLM Data Abstraction Systems}
% \title[Challenges in Specifying LLM-Based Data Abstraction Tasks for Clinical Research]{``I Know Where to Look'': Challenges in Specifying LLM-Based Data Abstraction Tasks for Clinical Research}
\title[Charting the Gaps Between Clinical Expert Needs and Unstructured Data Abstraction Tools]{``I Know Where to Look,'' But Does the LLM? Charting the Gaps Between Clinical Expert Needs and Unstructured Data Abstraction Tools}

\author{Venkatesh Sivaraman}
\email{venkatesh.sivaraman@ucsf.edu}
\author{Rigney Turnham}
\author{George Bonano}
\affiliation{%
  \institution{University of California, San Francisco}
  \city{San Francisco}
  \state{CA}
  \country{USA}
}
\author{Nevin Aresh}
\affiliation{%
  \institution{Stanford University}
  \city{Palo Alto}
  \state{CA}
  \country{USA}
}

\author{Renumathy Dhanasekaran}
\authornote{Clinical collaborators listed in alphabetical order.}
\affiliation{%
  \institution{Stanford University}
  \city{Palo Alto}
  \state{CA}
  \country{USA}
}

\author{Margaret Guo}
\author{Sindhu Kubendran}
\author{Olivia Lin}
\author{Jonathan D Louie}
\author{Kristan Olazo}
\affiliation{%
  \institution{University of California, San Francisco}
  \city{San Francisco}
  \state{CA}
  \country{USA}
}

\author{Jeanne Shen}
\affiliation{%
  \institution{Stanford University}
  \city{Palo Alto}
  \state{CA}
  \country{USA}
}

\author{Harish Vasudevan}
\author{Jeanette Wong}
\affiliation{%
  \institution{University of California, San Francisco}
  \city{San Francisco}
  \state{CA}
  \country{USA}
}

\author{Emily Alsentzer}
\author{Jason A Fries}
\affiliation{%
  \institution{Stanford University}
  \city{Palo Alto}
  \state{CA}
  \country{USA}
}

\author{Anobel Odisho}
\author{John Gordan}
\author{Jean Feng}
\author{Julian C Hong}
\correspondingauthor
\affiliation{%
  \institution{University of California, San Francisco}
  \city{San Francisco}
  \state{CA}
  \country{USA}
}
\affiliation{%
  \institution{Weill Cancer Hub West}
  \city{San Francisco}
  \state{CA}
  \country{USA}
}
\email{julian.hong@ucsf.edu}
\correspondingauthor

\renewcommand{\shortauthors}{Sivaraman et al.}

\begin{abstract}
Clinical data abstraction, the process of distilling structured information from patient records, plays a key role in advancing knowledge about diseases such as cancer.
Information extraction (IE) with large language models (LLMs) could accelerate this process, but it is unclear whether current frameworks effectively support clinical researchers without AI expertise. 
  To address this, we co-designed an interactive LLM-based abstraction system called Libretto with seven cancer research teams, then evaluated the system's ability to help them answer real-world research questions.
  We found that while clinicians knew where and how to annotate complex concepts in patient notes, in twelve of fourteen tasks they faced barriers to replicating those intuitions with LLMs.
  Contextual note reliability judgments, difficulties in steering vibe-coded prompts, and inflexible evaluation strategies necessitated fundamental changes to the IE workflow.
  Our results highlight open problems for HCI research to bridge the gaps between AI data work tools and clinical users' needs.

    % To study this human-computer interface, we designed an interactive system called Libretto in collaboration with seven cancer research teams, which initially reflected current frameworks and evolved over the course of the study to address the needs/challenges expressed by the research teams.

  % We addressed this challenge by designing an interactive system called Libretto in collaboration with seven cancer research teams, who posed abstraction tasks often surpassed prior (medical?) data abstraction benchmark tasks in complexity (and nuance?).
  % We evaluated Libretto's ability to support these researchers' abstraction tasks, which often surpassed prior (medical?) data abstraction benchmark tasks in complexity (and nuance?).
  % We identified unforeseen barriers that made even seemingly simple tasks difficult to specify, including implicit conventions about which notes to search in and in what order, difficulty steering vibe-coded prompts, and inflexible evaluation strategies. 
  % Our results establish novel research directions for tools that bridge the gaps between AI tools for data work and expert users' needs.
    % We designed a functional data abstraction system, called Libretto, then  as they used the tool to advance their projects.

\end{abstract}

\maketitle

\section{Introduction}

\begin{figure*}
 \includegraphics[width=\textwidth]{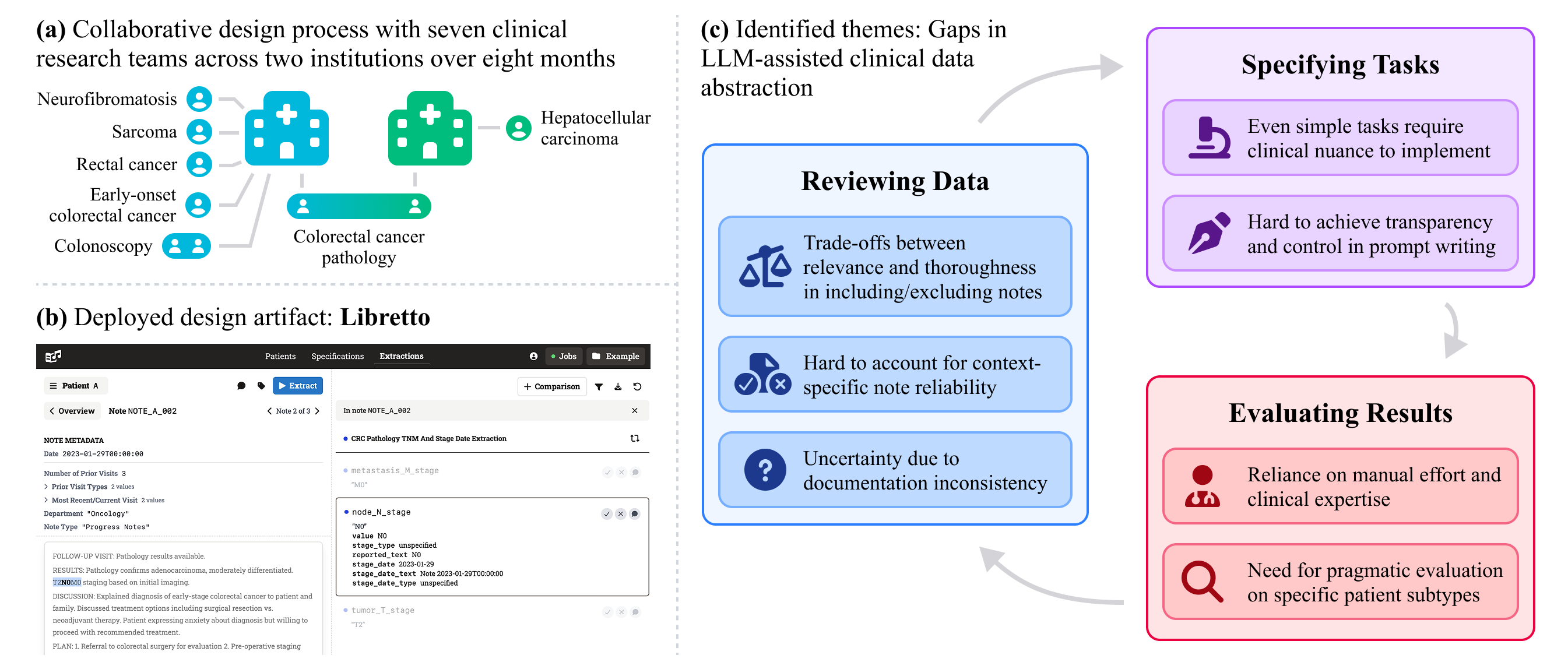}
  \caption{(a) This work documents a longitudinal co-design effort with seven cancer research teams, with the goal of developing a system to abstract structured fields from unstructured patient notes. (b) The resulting system, called Libretto, allowed clinician researchers to interactively specify, refine, and spot-check abstraction tasks. (c) Using this design artifact, clinician researchers took on more complex abstraction tasks than previously possible, but at the same time identified gaps between their intuitive judgment and the current mechanics of LLM-based abstraction.}
  % \Description{Enjoying the baseball game from the third-base
  % seats. Ichiro Suzuki preparing to bat.}
  \label{fig:theme-summary}
\end{figure*}

Many advances in healthcare research are driven by knowledge obtained from patient records. 
For example, clinical researchers have improved our understanding of life-threatening conditions such as cancer by distilling patient trajectories into structured formats that can reveal quantitative patterns~\cite{lancet_cancer_2025}.
The success of these analyses depends on accurate, scalable \textit{clinical data abstraction}~\cite{watzlaf_clinical_2021}, or the transformation of dense patient records (often including unstructured data such as clinical notes and images) into clean, rigorously-defined tabular fields.
However, clinical data abstraction has long been an expensive, manual, and error-prone process, even when technology is involved~\cite{miller_automated_2024,zozus_training_2019,liu_using_2026}.
% At its core, clinical data abstraction requires interpreting clinical data originally collected for a different purpose (e.g., patient care) for a new purpose (e.g., research), which .
Interactive systems that can streamline clinical data abstraction therefore have the potential to foster more expedient, reliable discoveries and improve patient care.

Recently, large language models (LLMs) have shown strong potential to accelerate clinical data abstraction by automating information extraction (IE) from unstructured patient data.
One study estimated over 30-fold time savings using an LLM compared to fully manual review for extracting pathology information, with the LLM outputs often attaining high accuracy~\cite{choi_developing_2023}.
LLM technologies have also enabled medical informatics researchers to explore not just the simple entities and relations extracted by classical IE tools~\cite{fu_clinical_2020}, but also more open-ended, clinically meaningful tasks such as quality improvement~\cite{williams_using_2026} and matching of patients to clinical trials~\cite{wornow_2025_zero}.
However, adoption of these technologies in the wider field of clinical research so far remains limited.
Existing methods for running LLMs across large clinical datasets require technical knowledge about database infrastructure, programming, and prompt engineering~\cite{hein_iterative_2025,choi_developing_2023,hu_improving_2024}, which can be difficult to navigate even with the assistance of AI coding agents.
Additionally, as with many open-ended research tasks, investigators often do not bring a well-defined question and instead need to see preliminary abstraction results to identify critical nuances~\cite{subramonyam_bridging_2024,sivaraman_tempo_2025}.
% The specification of an abstraction task is akin to problem formulation in predictive modeling and risk assessment, where the choices of data cohort, target variable, and covariates influence the quality of the final results at least as much the optimization of the model itself~\cite{feng_human-ai_2026,bhattacharya_exmos_2024}.
% Domain expertise during these early stages has proven critical to ensure that the resulting predictive model produces useful outputs~\cite{sivaraman_tempo_2025,passi_problem_2019}.
% Designing clinical data abstraction tasks requires expert clinical judgment and iterative cycles of manual review, and is sometimes considered as much an art as a science~\cite{allison_art_2000}.
Even if artificial intelligence (AI) can optimize the information extraction process itself, the questions of for whom, what, and how to extract information remain challenging to answer.
Therefore, it is unclear how clinical researchers, especially those without technical expertise, can best work with LLM-based data abstraction tools to produce valid, meaningful outputs.

% As the volume of readily available clinical data has grown in recent years, clinician researchers have become increasingly interested in abstracting structured information for retrospective analyses to understand patient prognoses, effectiveness of treatments, and gaps in healthcare delivery~\cite{zhao_changing_2022}.
% At the same time, LLM technologies have prompted a shift from simple extraction of entities and relations~\cite{fu_clinical_2020} to more open-ended, clinically meaningful tasks such as quality improvement~\cite{williams_using_2026} and patient-trial matching~\cite{wornow_2025_zero}.
% Therefore, in this work, we aimed to close the loop of clinical data abstraction by allowing clinician researchers to specify, refine, and evaluate their \textit{own} abstraction tasks.
% We therefore envisioned clinicians using LLMs for chart abstraction similarly to how scientists are using LLM agents to automate programming tasks~\cite{obrien_how_2025}.

The key contribution of this work is a longitudinal design research effort, using methods adapted from co-design~\cite{kleinsmann_barriers_2008,sedlmair_design_2012} and action design research~\cite{sein_action_2011}, to build and evaluate an interactive system for AI-assisted data abstraction.
We worked with seven research teams (9 direct collaborators, depicted in Figure \ref{fig:theme-summary}a) across two institutions, who each brought their own data abstraction goals related to the diagnosis and treatment of cancer.
We first designed and deployed a working prototype system called Libretto (Figure \ref{fig:theme-summary}b), which framed clinical data abstraction as a model formulation task in which researchers would iterate between reviewing source data, specifying tasks, and evaluating the results~\cite{passi_problem_2019,sivaraman_tempo_2025}.
We then qualitatively evaluated how well this system supported researchers as they defined fourteen different abstraction tasks across their respective projects.
Directly collaborating with these researchers provided a unique opportunity to study the evolution of clinical data abstraction problems in multiple contexts at once, improving the robustness of our findings compared to co-design within a single project~\cite{li_endoextract_2026,kothari_when_2026}.
We also observed LLMs' behavior on realistic tasks and real patient notes from our institutions' electronic health records (EHR), which are often messier and less consistent than those found in standard medical text datasets~\cite{rahman_generalization_2024}.
Finally, this study was conducted over the course of eight months spanning initial ideation to specification refinement, allowing us to observe real-world deployed tool use as participants became more familiar with LLM behaviors.
% In return for participants' time ideating and providing feedback on our system, we performed the necessary data science work behind the scenes so that participants could meet the goals of their respective projects.
% The prototype system developed throughout this design process is called Libretto.

% During the first phase of the co-design process, we designed and implemented a working system (called Libretto) based on the requirements expressed by participants and previous literature on clinical information extraction (IE).
% The resulting system structures clinician-AI collaborative data abstraction in a loop between three stages: Understanding Data, Specifying Tasks, and Evaluating Results.
% This system provided a functional interface for LLM-assisted chart abstraction with which to evaluate the challenges and limitations of a canonical workflow when used by clinicians.
% In the second phase, we evaluated whether the Libretto workflow could support participants' abilities to answer their research questions.
% Participants used the tool to create fourteen different ``specifications'' for various abstraction tasks, both independently and in periodic meetings with the research team.
% Through these sessions, we helped improve participants' patient cohorts and prompts, iteratively refined the system, and generated reusable design insights for how LLM-powered systems could better meet researchers' needs.
Our study revealed that LLM data abstraction for \textit{real-world, clinically meaningful} tasks is far from simple, despite LLMs' strong performance in proof-of-concept tasks (see Figure \ref{fig:theme-summary}c for a summary of identified challenges).
While clinicians developed some specifications that were analogous to prior IE literature~\cite{choi_developing_2023,balasubramanian_leveraging_2025,kim_optimizing_2025}, many were either more complex than previously-explored tasks or required greater nuance.
Of the fourteen task specifications attempted, only two could be completed without significant iteration.
Much of this complexity stemmed from clinicians' domain-specific requirements around note reliability, uncertainty, temporality, and synthesis across multiple sources, all of which could influence the validity of results and required bespoke strategies to handle.
% Even extraction tasks that clinicians deemed straightforward ended up needing bespoke strategies to account for these contextual judgments that are not often addressed in the IE literature.
Moreover, these strategies were difficult for clinician researchers to communicate to the LLM despite having the ability to vibe-code prompts, indicating that the abstraction prompts were not as inherently interpretable or controllable as is often assumed.
Finally, the researchers expressed a need for ad-hoc evaluation of the outputs to avoid time-consuming manual labeling without sacrificing rigor.
The challenges identified in this work reveal important directions that HCI research can advance, bringing together interaction design and AI innovation to support clinically-relevant, accurate, and scalable data abstraction.

\section{Background and Related Work}

\subsection{Clinical Data Abstraction with Artificial Intelligence}

Our work focuses on the core task of clinical data abstraction (also called chart review), in which researchers, who are often also clinicians, distill structured information from electronic health records (EHRs)~\cite{allison_art_2000,watzlaf_clinical_2021}.
Clinical data abstraction is challenging and time-consuming because EHR data is typically highly heterogeneous and collected for a different purpose than research (i.e., billing and patient care), so clinically important values may be at best inconsistently or implicitly described~\cite{rosenbloom_data_2011}.
Therefore, the medical informatics community has long been interested in developing automated information extraction (IE) systems to support data abstraction.
``Classical'' tasks for clinical IE include concept extraction and named entity recognition~\cite{fu_clinical_2020}, temporal understanding~\cite{sun_temporal_2013}, and summarization~\cite{zhang_leveraging_2021}.
Well-established methods for these tasks (e.g., MetaMap~\cite{aronson_overview_2010} and cTAKES~\cite{savova_mayo_2010}) allowed larger-scale analysis than clinician researchers could undertake alone, yet they were limited in the depth of information they could capture from each record due to the heterogeneity of text phrasing, lack of contextual understanding, and high volume of irrelevant text~\cite{tamine_semantic_2021,seifaddini_information_2026}.

Large language models (LLMs) have largely surmounted all of these limitations, demonstrating high accuracy on a range of clinical IE tasks including concept extraction~\cite{agrawal_large_2022,balasubramanian_leveraging_2025,liu_human_2025}, note summarization~\cite{van_veen_adapted_2024}, and even more subjective tasks like identifying barriers to optimal care~\cite{vossler_fuzzy_2026}.
These promising results suggest not only that LLMs can effectively augment clinical researchers' ability to perform existing IE tasks, but also that it can enable entirely new research questions.
However, the process in most abstraction workflows is still predominantly manual~\cite{watzlaf_clinical_2021}, potentially because it is still difficult and time-consuming to achieve good results for clinically meaningful questions.
In cancer informatics, for example, recent studies have used deep learning methods and LLMs to extract tumor staging~\cite{choi_developing_2023}, metastasis (cancer spread beyond the original site)~\cite{soysal_identifying_2017}, and recurrence after treatment~\cite{bayona_large_2026}.
However, these methods all required significant note pre-processing and task simplification to achieve high accuracy, and a similar study conducted with minimal pre-processing found a higher error rate due to note ambiguity and conflicting information~\cite{balaji_metastasis_2026}.
As \citeauthor{allison_art_2000} describe it, chart abstraction resists ``a `cookbook' approach'' and demands ad-hoc clinical judgment~\cite{allison_art_2000}, whereas LLMs typically require well-defined instructions~\cite{hu_improving_2024}.
These prior findings emphasize the need to help clinical researchers define tasks that meet their contextual requirements while maintaining LLM accuracy.

    % 1. What is chart abstraction?
    % 2. Lots of methods for clinical information extraction (cTAKES, MetaMap, …, deep learning, LLMs)
    % 3. These methods typically assume the problem formulation is fixed and objective
    % 4. However, real-world chart abstraction has long been known to be a highly iterative process (The Art and Science of Chart Review). We need to be able to support this iteration

Another barrier to integrating LLMs into clinical data abstraction is that running LLMs across large datasets currently still requires technical expertise.
For example, Google's Healthcare Natural Language tools and others from industry and academia~\cite{goel_introducing_2025,burdenko_medical_2024,hsu_llm-ie_2025,shankar_docetl_2025} provide command-line and web API-based interfaces to run LLMs on clinical data at scale.
These systems allow data scientists to create and refine sophisticated pipelines, but they assume knowledge of EHR data infrastructure, coding, and prompt engineering. 
Although AI coding agents can support this process, it is difficult for non-technical users to validate the results, especially to the level needed for scientific publication~\cite{obrien_how_2025,drosos_its_2024}.
As a result, these technical systems require close collaboration between clinical experts and data scientists, which can lead to long feedback loops and redundant effort without a carefully-designed process~\cite{kothari_when_2026,passi_problem_2019}.
%has a predefined, clear-cut, and objective task definition already.
% As a result, clinician researchers and data scientists have limited ability to collaborate on how data should be abstracted, leading to long feedback loops and often redundant effort~\cite{kothari_when_2026}.

\subsection{Human-AI Interaction for Medical Experts}

% The design of AI-powered interactive systems for healthcare has been a topic of great interest in HCI, given the growing recognition that highly accurate AI does not necessarily translate to effective adoption and improved outcomes~\cite{andersen_introduction_2023}.
Our work builds on extensive HCI and healthcare literature on using AI to help medical providers deliver care, including clinical decision support~\cite{mastrianni_recommend_2025,sivaraman_intelligent_2026}, workflow improvements~\cite{sendak_human_2020}, and patient and caregiver engagement~\cite{klein_approaching_2026,solano-kamaiko_sharing_2026}.
This line of work has repeatedly shown the importance of tailoring AI outputs to fit specific information needs~\cite{sivaraman_intelligent_2026,zhang_rethinking_2024}, accounting for varying levels of confidence in different information sources~\cite{zakreuskaya_managing_2026,kaltenhauser_you_2020}, and leaving space for decision-makers to ideate and deliberate~\cite{everett_tool_2026}.
We find that many of these themes translate to clinical researchers' use of AI for data abstraction as well. 

Our work also draws on literature in which AI is used to support \textit{data analysis} by non-technical users, including clinicians.
Recent work has highlighted the potential of AI tools to support researchers in writing data analysis code, particularly given the complexity of modern EHR databases~\cite{ma_tempoql_2025,obrien_how_2025,sun_lambda_2026}.
However, clinical researchers often still face challenges in AI-assisted data analysis.
In particular, it is often difficult for non-technical users to articulate the right task without iterative development~\cite{kothari_when_2026}, a problem called the ``gulf of envisionment''~\cite{subramonyam_bridging_2024}.
Related literature in the development of predictive models~\cite{feng_human-ai_2026,sivaraman_tempo_2025} and other types of data analysis code~\cite{gu_how_2024,drosos_its_2024} suggests that output transparency, rapid prototyping, and knowledge sharing with more technical users are critical design components to bridging this gulf~\cite{kothari_when_2026}; we aimed to embody these principles in our design artifact.

% The two streams of research within this broad field that are most relevant to our work are (1) the use of AI to improve direct patient care, and (2) AI tools for data analysis, of which healthcare data is a special case.

    % 1. Two main streams of work: (1) how AI outputs can support clinical decision-making and reasoning (reasoning cues, human-centered explainable AI, AI for deliberation, etc.), and (2) how AI agents can support end-user programming and data science with domain experts (interplay on developing problem specifications and evaluating results, tools for thought, etc.)
    % 2. Chart abstraction bridges these two lines of research as a task requiring both iteration between multiple types of expertise, and clinical reasoning and ad-hoc judgment

So far, human-computer interaction (HCI) literature addressing abstraction from patient notes has been somewhat limited, and clinician-facing data abstraction tools such as REDCap are predominantly manual~\cite{harris_research_2009}.
Most HCI efforts around clinical text have focused on how providers write and review patient notes at the point of care, integrating capabilities such as decision support, summarization, scribing, and exploratory visualization~\cite{sultanum_more_2018,li_automating_2021,li_exploring_2026}.
However, while much of this work emphasizes adapting to and understanding the nuances of an individual patient~\cite{li_exploring_2026}, the need for clinical researchers to analyze large sets of patients while retaining this nuance remains under-supported.
Recently, \citet{li_endoextract_2026} developed an interactive system for clinical researchers to review the outputs of an LLM-based IE pipeline in endometriosis ultrasound reports.
The Evaluate Results stage of our design artifact draws inspiration from their work; however, their system does not support the iterative process of \textit{specifying} the tasks that the LLM should extract.
Meanwhile, our work builds on a recent line of research studying the applications of LLMs in oncology~\cite{li_exploring_2026,verma_rethinking_2023}, where patient care and research are often intertwined due to the ongoing pursuit of better diagnosis and treatment practices~\cite{verma_rethinking_2023}.
Unlike these prior works, our study design focused on helping clinicians define their own LLM-based abstraction tasks rather than interacting with LLM outputs as an end user, allowing our system to generalize broadly to future data abstraction tasks.
% [HCI pre-LLM systems]

\section{Methods}

We aimed to understand how current LLM-based workflows would support or hinder data abstraction efforts through a longitudinal, collaborative design process with clinician researchers.
Our method can be considered a form of co-design, an iterative process by which users, developers, and other stakeholders build shared understanding towards the common goal of realizing a designed object~\cite{sedlmair_design_2012,kleinsmann_barriers_2008}.
Our approach is also related to action design research (ADR), a hybrid of action research and research through design that uses collaboration with target users and direct intervention in their work environment as tools to produce design knowledge~\cite{sein_action_2011,zimmerman_research_2007}.
ADR emphasizes real-world problem solving as the setting for knowledge creation, the use of theory-driven artifacts, and embedding evaluations throughout the design process, all principles followed in our study.

% The study comprised two phases: open-ended iterative design from early prototypes resulting in initial requirements, followed by the development of a functional design artifact called Libretto and in-depth usage and feedback sessions. % set up for participants' actual research goals.
% We developed a working system because it is often challenging for people to evaluate low-fidelity prototypes of intelligent interfaces, since the capabilities and limitations of an AI system may not be evident until it is seen working on real-world data~\cite{yang_re-examining_2020}.
% Moreover, we wanted to allow the clinician researchers to work with the patient populations and abstraction tasks they were most familiar with, allowing them to provide nuanced and realistic feedback on barriers they encountered.
% This project therefore represents a combination of both highly contextually-specific inquiry and triangulation across users that is otherwise difficult to attain.

% Why co-design with functional prototypes? Most clinicians don’t have experience with LLMs or intuitively know how they will behave. Conducting the research in parallel with participants’ actual work allowed us to get realistic results (with participants who were motivated to see the project succeed)

\subsection{Research Setting and Clinical Collaborators}

This project took place within a larger, cross-institutional research effort to map the illness trajectories, treatment practices, and outcomes of patients with cancer. % by extracting information from unstructured notes.
Cancer is a highly heterogeneous family of diseases, and cancer diagnosis and treatment for an individual patient requires multi-factorial analysis of data across complex modalities. 
Cancer treatment decisions are usually based on outcomes derived from population averages; yet these predictions fall short for individual patients, leading to misdirected treatments and suboptimal outcomes~\cite{ramos-casallas_performance_2025}. 
As the volume of data available on cancer patient journeys and outcomes has grown, there is a critical need for tools that surface this knowledge to inform clinical decision-making~\cite{didier_application_2024}.  %As understanding of cancer has improved, the complexity of clinical decision-making can stretch the capabilities of expert physicians/
Many valuable clues to disease progression and response to treatment in cancer lie in unstructured notes such as pathology reports (analyses of biopsies and other specimens of cancerous tissue), imaging results, and assessments and messages by other medical providers~\cite{seifaddini_information_2026}.
However, reliably extracting these signals currently requires extensive manual validation for each type of cancer on which they are applied~\cite{balasubramanian_leveraging_2025,choi_developing_2023}.
By bringing together several clinical collaborators to co-design a chart abstraction system while defining their own tasks, we aimed to understand common needs across cancer types that would likely generalize to other clinical research domains.

\begin{figure*}
    \centering
    \includegraphics[width=\linewidth]{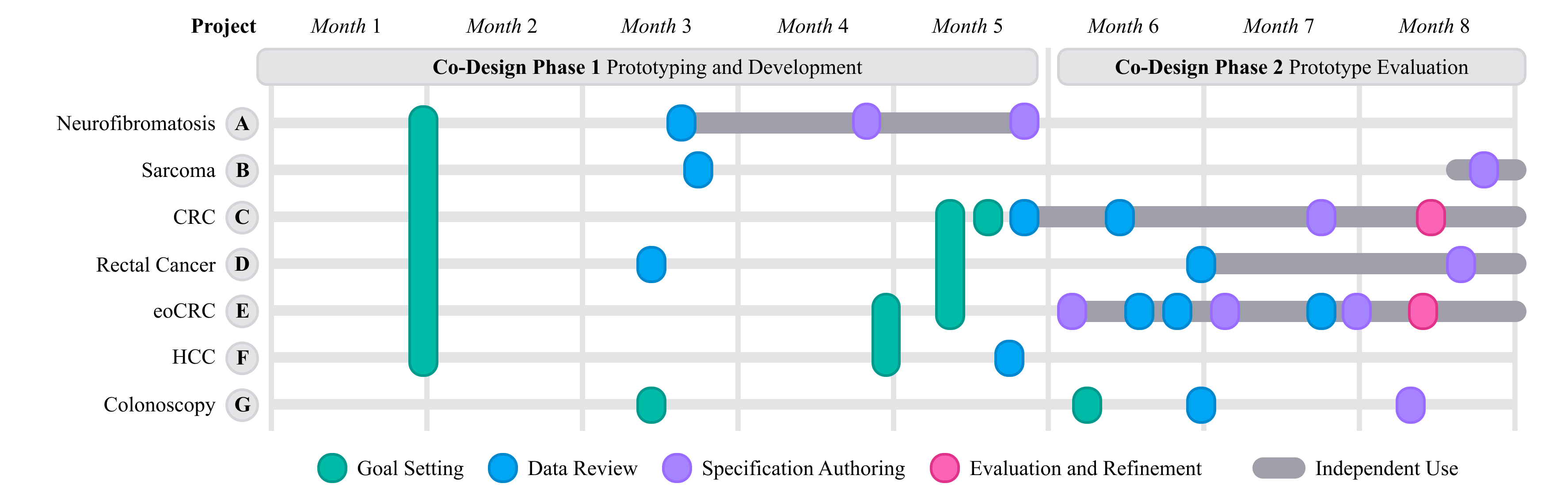}
    \caption{Timelines of involvement of each collaborator project throughout the two phases of co-design. Dark gray lines indicate periods in which clinical collaborators were using Libretto independently to create and edit specifications. Note that Projects A and F participated in the first phase but not the second due to scheduling and data access constraints, respectively.}
    \label{fig:project-timeline}
\end{figure*}

We established working relationships with nine researchers from seven different teams to develop clinical data abstraction tasks.
Details on each project are shown in Table \ref{tab:project-overview}, while the timeline of engagement in each project is shown in Figure \ref{fig:project-timeline}.
(In addition to the nine direct collaborators, several other researchers and assistants affiliated with each project joined these meetings but did not directly use our system.)
The projects spanned several cancer types, including cancers of the intestine (projects C, D, E, and G), the liver (project F), and the nervous system (project A), as well as rare cancers of the soft and connective tissues (project B).
Most participating researchers were affiliated with the institution of the core team, a large US-based academic health system.
Two additional members were recruited from a second academic institution and healthcare system to examine differences in abstraction workflows in another environment.
% We recruited the researchers from departments adjacent to those of the core team members within our institutions; therefore, we were able to capture expertise in several different cancer domains.

Since we supported these individuals in performing work relevant to their own research during our co-design process, we considered them to be both collaborators and study participants.
Those who directly used our system and contributed feedback were invited to be co-authors.
Throughout the paper we will refer to these individuals interchangeably as ``participants,'' ``clinical researchers,'' or ``collaborators.''
The core team, which led the co-design process, consisted of researchers in human-computer interaction, oncology, AI, and biostatistics.
We also worked with machine learning and data infrastructure engineers to extract the datasets needed for each collaborator's project and securely host our data abstraction platform while protecting patient privacy.

    % Each project will be assigned a code letter A-G, and participants will be numbered within each project. Only participants in the teams who actually used the tool themselves are listed
    % A - neurofibromatosis
    %     1 - Harish
    %     2 - Harish's student? Whoever used the tool
    % B - sarcoma
    %     1 - Margaret
    % C - colorectal cancer pathology
    %     1 - Jeanne
    %     2 - Jonathan
    % D - rectal cancer provider rationale
    %     1 - Sindhu
    % E - early onset colorectal cancer 
    %     1 - Olivia
    % F - HCC
    %     1 - Renu
    % G - colonoscopy screening
    %     1 - Kristan
    %     2 - Jeanette

\begin{table}[]
    \centering
    \small
    \begin{tabular}{lp{2cm}p{3.5cm}p{1.75cm}p{1.75cm}p{2cm}}
    \toprule
\textbf{Project Code} & \textbf{Focus Area} & \textbf{Primary Research Question} & \textbf{Participant (Institution)} & \textbf{Role} & \textbf{Size of Data Pull (\# Patients)} \\ \midrule
A & Neurofibromatosis & What factors increase the likelihood of developing a malignant tumor? & A1 (primary) & Attending & 4,157 \\
B & Sarcoma & How do tumor genetics affect prognosis and response to treatment? & B1 (primary) & Fellow & 9,375 \\
C & Colorectal Cancer (CRC) & How can we predict cancer recurrence given pathology and imaging information? & C1 (secondary) & Attending & 1,902 \\
 &  &  & C2 (primary) & Fellow & 3,700 \\
D & Rectal Cancer & What provider decision-making factors influence the decision to recommend adjuvant therapy? & D1 (primary) & Fellow & 6,961 \\
E & Early-Onset \newline Colorectal \newline Cancer (eoCRC) & What factors in initial diagnosis and treatment influence the chance of survival? & E1 (primary) & Fellow & 1,960 \\
F & Hepatocellular Carcinoma (HCC) & What factors influence cancer recurrence after surgical tumor removal? & F1 (secondary) & Attending & (not extracted) \\
G & Colonoscopy & What factors explain why patients do not receive a colonoscopy to screen for cancer after an abnormal fecal immunochemical test (FIT)? & G1 (primary) & Clinical Data  \newline Analyst & 1,112 \\
 &  &  & G2 (primary) & Clinical Data \newline Analyst & (same as above) \\  \bottomrule
\end{tabular}
    \caption{Overview of clinical collaborators and projects in this study. Attendings are senior clinical research faculty, while fellows are trainees with full medical training.}
    \label{tab:project-overview}
\end{table}

\subsection{Clinical Data Retrieval}
During the early stages of each project, we worked with the researchers to define the cohort of patients and notes they were interested in.
At the primary institution, we obtained IRB approval to access fully-identified patient data and notes from the Epic Clarity and Caboodle databases, which are derived from the Epic EHR and formatted for reporting and analytics.
While this institution also provided de-identified notes, the identified notes were generally seen as more reliable because the machine redaction often removed useful information such as landmark dates.
In contrast, at the secondary institution we were unable to access identified notes and instead relied on the collaborators bringing their own data in CSV format.

For most projects, patients were included on the basis of having at least two visits to relevant departments (e.g., medical, surgical, or radiation oncology) and a diagnosis code related to the disease of interest.
This resulted in cohorts of between 1,000 and 10,000 patients per project.
Clinical notes, imaging reports, and pathology reports were included based on the type of note and the type and department of the encounter that generated them, yielding anywhere between 1 and 500 notes per patient.
This diversity in dataset size allowed us to observe the data abstraction workflow at a range of scales.

\subsection{Study Phases}

\begin{figure*}
    \centering
    \includegraphics[width=\linewidth]{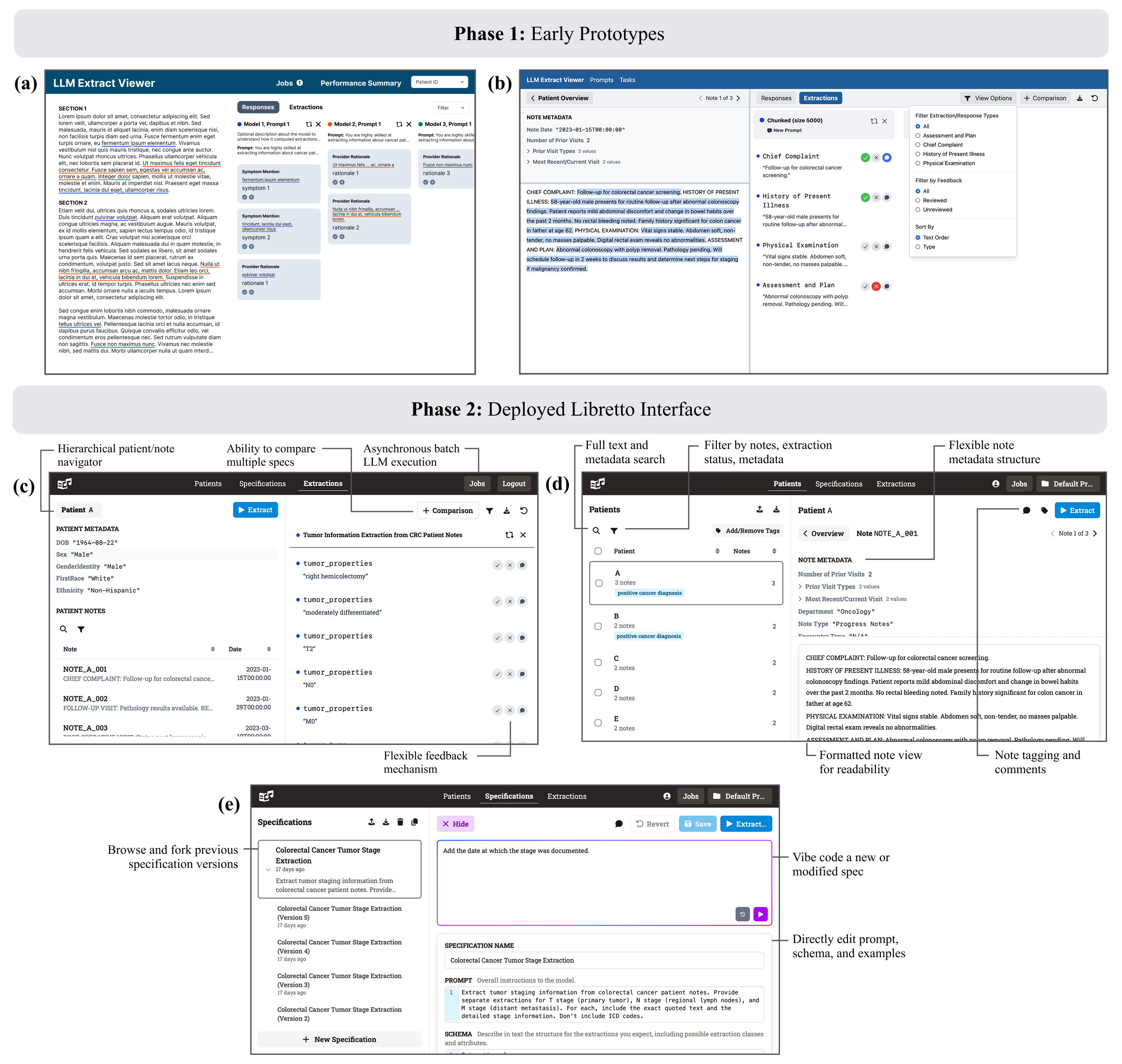}
    \caption{Representative screenshots of prototypes from the earliest stages of co-design to the working version of Libretto. (a) represents a non-functional mockup used to elicit initial feedback, while (b) shows the first version of the working system, with limited ability to review patients or define specifications. The final deployed version contained (c) an Extractions view focused on evaluating a single spec on one patient at a time for simplicity, with additional views for (d) browsing patients and (e) editing specifications.}
    \label{fig:interfaces}
\end{figure*}

The study proceeded in two phases, the first using early mockups and functioning prototypes, and the second using a stable version of Libretto.
The first stage, which took approximately five months, focused on establishing the goals of each collaborator's project and defining the design requirements for Libretto.
We held both large-group meetings across teams and smaller meetings with individual researchers to brainstorm ideas for the system design.
These meetings were loosely organized around a set of interview questions that probed (1) participants' level of experience with programming and AI-assisted clinical data abstraction, (2) how extraction from unstructured notes could support their goals, and (3) how they would prefer to evaluate the extracted data.
We also showed participants our preliminary designs and prototypes of the Libretto interface, some of which are shown in Fig. \ref{fig:interfaces}, and requested feedback on what capabilities it should include and how it could be most intuitive for clinicians.
These early meetings yielded valuable insights, particularly around the importance of selecting the right patients and notes and being able to review them within the interface before specifying any LLM tasks.
Using these insights, we iteratively formalized Libretto's conceptual workflow, developed the interface, and resolved usability issues.

In the second stage of the study, lasting about three months, participants used a stable, deployed version of Libretto to make progress on their data abstraction goals.
We aimed to meet with participants one-on-one as schedules allowed to develop abstraction specifications in Libretto and evaluate the results.
These meetings did not have a specific interview script; instead, we exchanged progress updates, addressed any of the participants' questions or challenges, and asked for feedback on the system.
In some cases, we also presented alternative designs based on participants' suggestions and elicited their opinions on if they would be more helpful than the current interface.

As shown in Figure \ref{fig:project-timeline}, each project also moved through its own stages of work within these larger phases.
We began with \textbf{Goal Setting}, where collaborators shared the context of their projects and we established the initial cohort definition and research question that we would address.
Then, we performed an initial data pull and showed it to the clinicians for \textbf{Data Review}, which often resulted in updates to the cohort definition.
In the next stage, \textbf{Specification Authoring}, participants used Libretto to define abstraction tasks.
Finally, participants conducted \textbf{Evaluation and Refinement} by inspecting the results of the specifications and making changes.
Note that these stages were sometimes combined within a single session or followed out of order; for example, we returned to Data Review after the first round of Specification Authoring in project E because the specification revealed that a different set of note types was necessary.

\subsection{Analysis}

Feedback and perspectives collected throughout all 28 sessions were analyzed using inductive thematic analysis.
With participants' permission, the sessions were video-recorded and transcribed using Zoom.
The transcripts were annotated using open coding by two members of the core team, who discussed and resolved any disagreements as they arose.
The resulting 636 codes were then clustered using affinity diagramming, which yielded 71 intermediate-level codes and nine high-level themes that are discussed in Section \ref{sec:results}.
    
\section{Design Artifact: Libretto}

The product of the first phase of design was a web-based clinical data abstraction system called Libretto (a reference to the often obscure yet meaningful texts of opera music, analogous to clinical notes).
The key insight in Libretto's design was to consider abstraction specification as an \textit{iterative, open-ended model formulation task}.
While classical IE systems had relatively fixed capabilities (i.e., mining a constant set of concepts), LLM-based systems are much more flexible, making it possible to define more meaningful tasks but also creating the possibility of misalignment with clinical needs.
These issues have been well-documented in HCI and ML literature on designing predictive models~\cite{passi_problem_2019,sivaraman_tempo_2025,feng_human-ai_2026,bhattacharya_exmos_2024}, giving us a conceptual lens to design the system and understand the challenges it induced.
The model formulation framing also aligns with the clinical understanding that chart review is as much an art as a science, requiring iteration and clinical judgment to achieve validity~\cite{allison_art_2000}.

We therefore designed Libretto to embed standard LLM-based IE features within an iterative development workflow with three stages: \textbf{Reviewing Data}, \textbf{Specifying Tasks}, and \textbf{Evaluating Results}.
This cycle was distilled from models of data science work, such as \citeauthor{tukey_exploratory_1970}'s model of exploratory data analysis~\cite{tukey_exploratory_1970}, \citeauthor{weick_organizing_2005}'s model of sense-making~\cite{weick_organizing_2005}, and recent studies of expert-in-the-loop predictive modeling~\cite{sivaraman_tempo_2025}.
In each stage, we designed functionality that aligned with clinicians' existing chart abstraction needs based on their early design feedback, as described below.
% The main features of Libretto are described below by how they support each stage.

\subsection{Reviewing Data} 
The importance of choosing the right set of patients and notes was apparent throughout our sessions, and participants consistently expressed the need to review the cohort we extracted and ensure that it matched their expectations.
This led us to develop the Patients view (shown in Figure \ref{fig:interfaces}d), which simply presented all patients in the data pull along with their associated notes.
One particularly important aspect of this page was its search and filter functionality.
Participants expressed many different search criteria, including for patients that they had cared for by medical record number (C1, D1), patients with particular types of notes (C1), patients with a specific diagnosis code (A1), or notes containing certain keywords (E1).
We implemented free-text search so that participants could search any of these fields in a consistent manner.

During the data pulls, we realized that each project would require slightly different structured metadata (e.g., the date or age of diagnosis or the patient's most recent date of follow-up), so we implemented support for fully flexible, searchable patient and note metadata in JSON format.
Participants also noted at first that reading the raw clinical text could be cumbersome because the source database did not store newlines or formatting information (see Figure \ref{fig:interfaces}b).
Therefore, we parsed the note text to add newlines and Markdown-style headings where possible.
Finally, users initially found it difficult to keep track of patients they had reviewed, so we allowed them to apply tags and freeform comments.
After implementing these features, participants could much more quickly assess the validity of a cohort and find relevant patients.

% \begin{figure}
%     \centering
%     \includegraphics[width=0.5\linewidth]{figures/patient_view.png}
%     \caption{The Patient view addressed researchers' needs around reviewing and verifying that the cohort they were specifying abstraction tasks on was valid, including (a) search, (b) filtering, (c) structured note metadata, (d) formatted note text display, and (e) tagging and comments to keep track of previously-reviewed patients.}
%     \label{fig:patient-view}
% \end{figure}

\subsection{Specifying Tasks} 
In this stage, participants would define abstraction ``specifications'' (specs) that could be run across a set of notes.
We built the Specifications view (Figure \ref{fig:interfaces}e) to work with LangExtract~\cite{goel_introducing_2025}, an open-source IE toolkit that provides a standard spec structure and handles chunking of large documents, repeated parsing to improve robustness, and fuzzy matching to map extractions back to the source text.
Accordingly, the Specifications view allows users to author specs that consist of a prompt (an overall model instruction), a schema (free-text description of the fields that should be extracted), and examples (one or more synthetic input notes and the expected output).

We quickly learned that LangExtract specs were too involved and unintuitive for clinician researchers to write manually within the time constraints of a session.
They much preferred tools to help them formalize their big-picture spec ideas, leading us to add a Generate button to ``vibe code'' a spec.
This feature allowed users to either create new specs from scratch or edit existing ones by typing a free-text prompt, which was appended to a system prompt detailing the expected output format.
Like the extraction process itself, our implementation was agnostic to the particular LLM but used PHI-compliant Microsoft Azure deployments of GPT-5 mini~\footnote{\url{https://developers.openai.com/api/docs/models/gpt-5-mini}} throughout the design process.

As participants created and refined their specs, we identified that participants needed ways to track the specs' version history and return to previous versions (C1, E1).
We addressed this by listing previous versions of a spec as editable specs nested within the current version. 
By being able to directly select previous specs, users could then easily compare results from one version to another or branch specs to create multiple alternatives.

% \begin{figure}
%     \centering
%     \includegraphics[width=0.5\linewidth]{figures/specification_view.png}
%     \caption{The Specifications view allows users to define abstraction tasks in a free-text structure, with the ability to (a) browse and fork from previous versions, (b) vibe-code a spec using the ``Generate'' feature, or (c) directly edit the spec as used by LangExtract.}
%     \label{fig:specification-view}
% \end{figure}

\subsection{Evaluating Results} 
Finally, the system needed to help users review the results of running a specification, primarily by evaluating outputs one patient at a time.
The Extractions view (Figure \ref{fig:interfaces}c) supports this by showing details for a selected patient on the left and model outputs on the right.
As shown in Figure \ref{fig:interfaces}a and b, the ability to highlight phrases in the source notes that led to an extraction was seen as a priority from the very earliest mock-ups, and was widely considered extremely useful.
Following the first phase of co-design, we adopted a hierarchical navigator for the patient information so that clinicians could first browse the notes they wanted to review. 
Clicking into an extraction shows additional model-extracted details and jumps to the note in which the information was found.
We also learned that comparison across specs was less important than quickly assessing the results of the most recent spec, so we improved the display of a single spec's results.
We added the ability to provide feedback in the form of approval, rejection, or free-text comments on individual extractions; these annotations could be used to filter exported results or quantitatively validate specifications.
Finally, we implemented the ability to download results as a CSV file so that researchers could easily load LLM-generated attributes into their existing research databases.

% \begin{figure}
%     \centering
%     \includegraphics[width=0.5\linewidth]{figures/extraction_view.png}
%     \caption{The Extractions view displays the results of LLM-based IE on selected patients and notes, with patient data on the left and LLM outputs on the right. It also allows for comparing specification outputs and providing human feedback to be saved alongside the results.}
%     \label{fig:extraction-view}
% \end{figure}

\subsection{Implementation Details}
Given the strong data security requirements in our healthcare institutions, the technical architecture of Libretto was designed for quick setup within diverse computational environments to minimize the approvals needed for a shared deployment.
The software was implemented as a web-based application using a FastAPI backend, Svelte frontend, and SQLite database, with both Docker and traditional command-line based deployment options for flexibility.
We connected Libretto to LLM services through our institutions' internal OpenAI and Anthropic API endpoints, which were approved for use with protected health information (PHI).
The system itself is agnostic to the LLM provider and can be connected to any provider supported by the LiteLLM open-source package,\footnote{\url{https://github.com/BerriAI/litellm}} including VLLM and Ollama for locally-hosted open models.
The system is open-source for development and adoption by the clinical research community.\footnote{GitHub link omitted for anonymity.}
    % 1. Define the iterative cycle: Understanding Data -> Specifying Tasks -> Evaluating Results
    %     1. Mirrors what is currently done with medical students
    % 2. editable Patient browser with tagging and search, 
    % 3. Specifications view with LangExtract, vibe coding
    % 4. Extractions view with live LLM runs, comparison, show in context

\section{Results}
\label{sec:results}

The clinician researchers in our study expressed excitement for the capabilities of Libretto and its LLM-driven workflow to help them take on previously-unanswerable research questions.
They were able to specify fourteen different abstraction tasks during the study.
For some of these tasks, such as identifying clinical factors (project A) or genetic markers (B, E), specifying the task was straightforward and resulted in plausible outputs after only one iteration.
However, in all but two cases, clinicians' task definitions revealed gaps between their intuitions and the mechanics of LLM-based data abstraction, which required significant iteration, AI expert intervention, or fundamentally different system capabilities to address.
The specific challenges and their resolutions for each task (numbered 1--14) are listed in Figure \ref{fig:prompt-challenges}.
% We also found previously-unforeseen sources of complexity that prevented the researchers from being able to formalize and refine their data abstraction tasks on their own.
Across all of the tasks, our qualitative analysis revealed nine major themes, which we detail below beginning with perceived sources of value of LLM-based data abstraction followed by challenges at each stage of the Libretto workflow.

% \begin{figure}
%     \centering
%     \includegraphics[width=\linewidth]{figures/theme_summary.png}
%     \caption{High-level themes identified from participants' use of Libretto in the study.}
%     \label{fig:theme-summary}
% \end{figure}

\subsection{Potential Benefits of LLM-Based Abstraction}

\subsubsection{Unlocking previously untapped information sources for research.} 
\begin{figure}
    \centering
    \includegraphics[width=\linewidth]{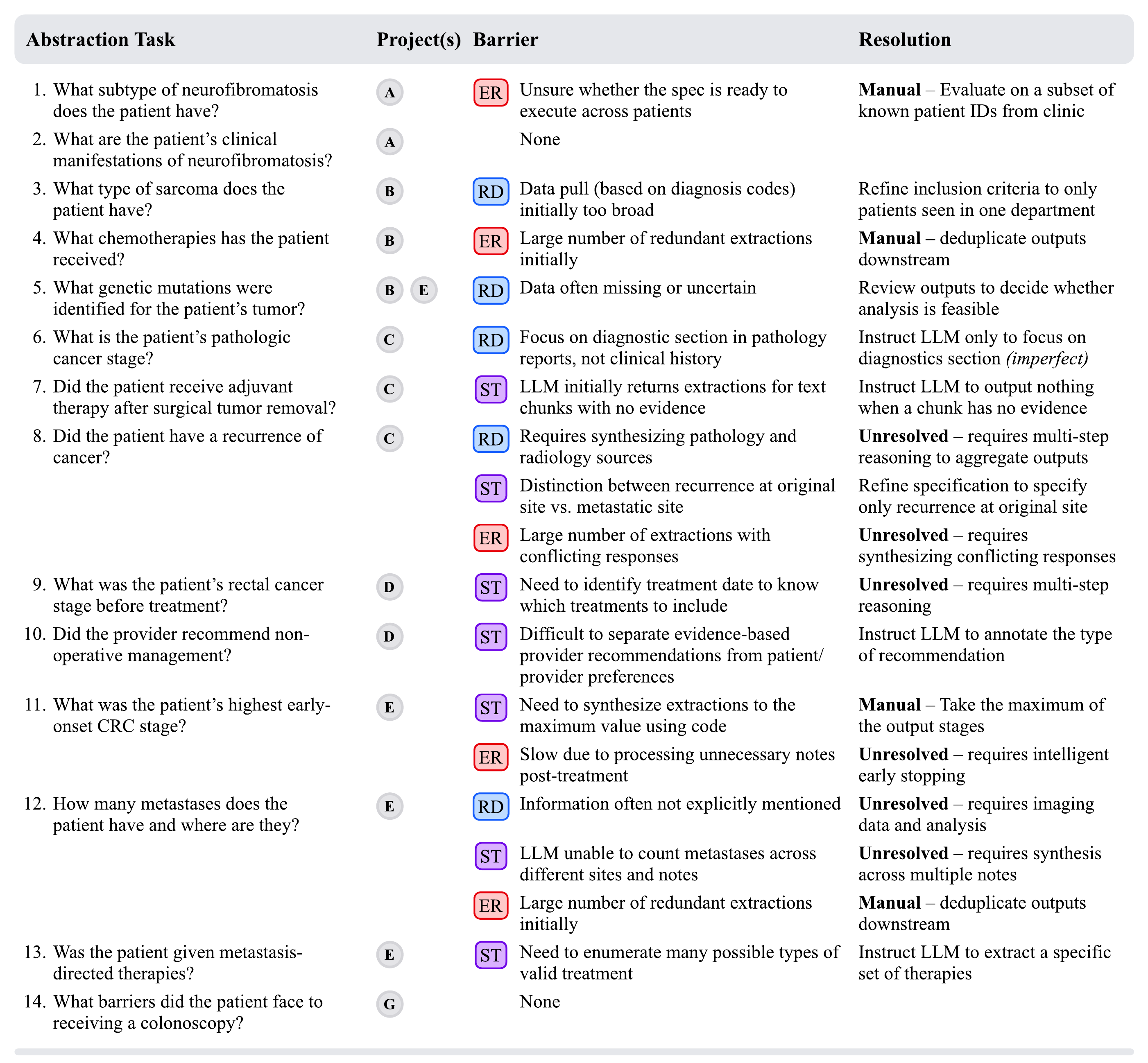}
    \caption{Summary of abstraction tasks that participants developed during the study, along with key barriers that were encountered during iterative development. Some barriers were resolved during the study within the Libretto workflow, while others required either manual code-level workarounds or more significant changes to the Libretto workflow to address. Barriers are categorized by the part of the workflow and thematic results they relate to, including Reviewing Data (RD), Specifying Tasks (ST), or Evaluating Results (ER).}
    \label{fig:prompt-challenges}
\end{figure}

As shown in Figure \ref{fig:prompt-challenges}, clinician researchers in our study brought specific abstraction goals at a wide range of difficulty, ranging from extracting formulaic responses out of a structured text-based form to inferring the reasons that a provider might have recommended a particular treatment.
They also described over-arching goals for LLM-enabled data abstraction that went beyond specific research needs, such as building a holistic longitudinal picture of patients for computational analysis or prospectively annotating patients as they are seen during care.
For example, A1 (researcher 1 in project A) described one of their goals as finding previously-uncaught signs of disease: \textit{``What I would like to understand is, in our cohort of folks [with a malignant tumor]... if we look back at imaging reports or look back at clinic notes, is there like a comment that this thing was already there and we just never saw it?''}
Other participants were particularly interested in augmenting their existing research with data sources that were conventionally not used in their domain, allowing them to make better predictions of survival or treatment response.
One aspect of Project C, for instance, was to learn how to incorporate radiology notes into a workflow that typically relies only on pathology to assess colorectal tumors.
Several participants were also interested in incorporating genetic information on patients' tumors, which was often stored heterogeneously across patients: \textit{``Some people like, snapshot, the [genetic report] PDF... some people only put the gene, some people put the gene and the actual mutation... So, they're all very different''} (B1).

Libretto was viewed as a promising way to address these long-standing questions which would otherwise be prohibitively difficult, all the more so because each abstraction task is only one step (e.g., a covariate or outcome variable) on the way to a larger analysis.
Project C, for example, required at least three different types of abstraction to characterize patients at baseline, capture their treatments, and detect whether they had a recurrence after treatment. 
As C2 described, \textit{``I've started to do this chart extraction manually... which, obviously, it all kind of adds up in terms of time.''}
% A1: Often when I'm seeing these patients, they would have had scans over the last 10 years, and we have to track the size of things... Usually I just pull up the imaging, but it would be nice if they could just even give me a summary from the reports, right?
% C2: Right now, like, treatment and prognostication is based pretty much on the, like, what pathology says, how they stage it, but what we're trying to do is, like, bring in what it… what the mass looks like on imaging as well, and kind of use that as another data point to try to better triangulate the course and the appropriate treatment.

\subsubsection{Enhancing validity and interpretability through custom structured information fields.} 
Given that many participants' goals centered around improving predictions of some kind, we probed as to why they could not simply use text-based deep learning models or existing tabular data from the patient's EHR.
However, participants believed it was both more technically and clinically useful to extract structured signals for use in downstream models: \textit{``Philosophically, my answer is your unstructured [model] is probably not good enough versus a noisy structured [model] that's derived from the unstructured [data]''} (D1).
These noisy structured signals were also perceived as more externally valid because they could be defined in a simple clinically-interpretable manner, whereas deep learning models could not: \textit{``Downstream ML or stats, it's very like straightforward... You don't want it to be fancy in the end''} (B1).
However, existing EHR structured data did not meet these researchers' needs because it was either too sparse or because the structured fields were not designed for retrospective research (B1, D1), leading them to explore abstraction from unstructured data.
Clinicians were generally familiar with using tools such as REDCap~\cite{harris_research_2009} to manually structure abstracted data (A1, C1, C2), and they expected these concepts to extend to LLM-based workflows.

\subsection{Reviewing Data: Note Reliability and Uncertainty}
\label{sec:reviewing-data}

Across the specifications we created, we generally found the LLMs we worked with to be effective at interpreting patient notes and following the extraction instructions it was provided.
However, this left more fundamental questions unresolved: Which notes should the LLM review? How should it prioritize information in the presence of conflicting or uncertain signals?
Probing clinicians' perspectives on these topics led to the following challenges.
% While our collaborators' attitudes toward LLM-based data abstraction were positive, in practice we faced many obstacles to making the process work, beginning with the patients and notes that were passed into the LLM in the first place.

\subsubsection{Trade-offs between relevance and thoroughness in including and excluding notes.}
In each of our collaborators' projects, we initialized Libretto with a dataset matching their desired patient inclusion criteria and a broad set of notes (potentially hundreds or thousands per patient).
For several collaborators, however, adding and removing notes to refine this data pull was a critically important step requiring time-consuming back-and-forth with data scientists on the team.
In many of the projects, the initial data pull contained a large portion of irrelevant notes for each patient, which caused the LLM to run more slowly and sometimes resulted in spurious outputs.
For example, E1 described their observations from reviewing the notes in Libretto's Patients view: \textit{``I probably will want to keep the notes from medical oncology, radiation oncology, and surgical oncology, but I noticed that a lot of patients have notes from when they're hospitalized... and [those] are not so helpful.''}
Since the clinical researchers were not familiar with how notes were structured and categorized in the underlying EHR database, these questions necessitated hours of discussion and iteration between technical and clinical team members to make sure the right notes were pulled.

Exclusion criteria could also be tricky to define using structured data alone.
Some described a tension between increasing the heterogeneity of their cohort and retaining potentially useful information when applying rule-based exclusions, such as by filtering out patients seen decades ago when treatment practices were different (D1, E1) or patients with related but distinct disease types (B1).
In the case of project E (which focused on early-onset colorectal cancer), researcher E1 wanted to include only patients with metastatic disease but found that this information could itself only be found in unstructured notes.
Therefore, we first created a specification to extract cancer stage from notes (Task 11), the results of which were used to create a new dataset in Libretto containing only those patients with stage IV cancer (representing metastatic disease).

\subsubsection{Difficulties accounting for context-specific note reliability judgments.}
Many researchers had previously reviewed patient records manually as part of research or clinical practice (A1, E1), and brought intuitions about which notes were likely to be most useful for their specific abstraction task.
For example, A1 mentioned that \textit{``diagnosis [date] is one that we define [by] pathologic specimens that have surgery, or date of resection. So we would never pull from the provider note because... that is not always accurate.''}
These less reliable notes were often still necessary to provide to the LLM as auxiliary context, but participants wanted ways to prevent the LLM from using them as a primary source.
In some cases even specific sections within a single note were viewed as less reliable, such as Diagnostics versus Clinical History in pathology reports for Task 6: \textit{``We've got two conflicting extractions from this one note... But the one that I would actually use for my project would be the [diagnostics section]... I could, you know, tell it, `okay, only pay attention to content in the diagnostics section, as opposed to the clinical history section,' but not all of the notes may be structured that way''} (C1).
Similarly, some participants noted that the most recent notes for a patient often contain all the necessary information (B1, E1), although prior notes would still be important if the most recent one contained erroneous ``copy-forward'' text (when providers transfer relevant content of previous notes into current notes, sometimes with slight modifications).

The Libretto Specification view initially offered the ability to define SQL queries for which patients and notes the LLM should review.
However, most users were not able to understand or edit these queries, and it was often difficult even for the technical team members to express clinicians' conceptions of note reliability in SQL.
Collaborators therefore suggested that the tool should err on the side of reviewing more notes, and present the outputs in a hierarchy of information quality that matched their intuitions (A1, C1, C2).
It could also be possible to combine programmatic filtering with the use of a smaller language model to semantically assess note reliability.
% C1: we've got two, sort of, conflicting extractions from this one note. But the one that I would actually use, right, for my project would be the, the other one, Pathology. [So you could update the spec to tell it to sort of only look at the pathology notes, or something like that.] How would it know that, though? Because that text was actually part of the pathology note. [Yeah, if… if it's not in the metadata, and not in the note, there's no way for the LLM to know what kind of note it is.] Okay. I mean, I think there will be, obviously some, you know, differences, right, in the way that nodes are structured, especially the older ones compared to the newer ones. I mean, I could, you know, tell it, okay, only pay attention to content in the diagnostics section, as opposed to that clinical history section, but not all of the notes may be structured that way to have a distinct, you know, separate clinical history section.

% 1. Deciding what notes to include and exclude is itself challenging: a naive data pull might have many irrelevant notes, yet these other note types are often helpful for triangulating information.

% 2. Participants make domain- and task-specific assessments of reliability for patients, notes and even sections within notes, often dependent on which types of providers and how the patients were seen. It’s challenging for them to articulate these hierarchies to the LLM.

\subsubsection{Uncertainty due to documentation inconsistency.}

Apart from varying reliability across different types of notes, the researchers were also concerned about the possibility of missingness and lack of explicit documentation for many concepts of interest.
They frequently offered clear intuitions about which types of information would be easier or harder to extract, often based on their own experience reading and writing patient notes.
For example, E1 described that colon cancer pathology would be easy to extract for Task 6 because \textit{``there are, like, 10 different things that they always are supposed to assess for''} in a semi-structured form.
On the other hand, variability across different providers was perceived as a primary obstacle to automated extraction, particularly for oncologist notes.
These notes document the process of deciding on a patient's overall care plan based on multiple sources of information, including pathology, imaging, and clinical history.
While sometimes these notes could be interpreted on their own because they contained entire quoted reports from other providers, other times they might only contain snippets or even mis-typed versions of the other notes (E1, B1).
Dates were considered particularly problematic, as they were needed to assess window-based criteria (e.g., recurrence within 3 years of initial treatment), but often written vaguely in the notes (e.g., ``six months ago'').
Another challenge was that these reports might often omit important decision-making details such as why a particular treatment was chosen, forcing researchers to \textit{``pick up on a vibe''} (D1) to assign a label.

Participants described these uncertainties as being largely inevitable in clinical data abstraction. 
However, although they were aware that the uncertainty existed, they lacked ways to externalize it or request it from the LLM.
As C1 put it, it would be \textit{``helpful for us to be able to just automatically flag which extractions are correct,''} something that could not be realistically labeled without intensive human effort.
Other participants suggested that the LLM quantify its uncertainty through confidence ratings or temporal ranges (A1, B1), though these tend to be unreliable~\cite{xiong_can_2024}.
The most realistic solution given the tools they had was to combine manual patient review and iterative prompt development to decide if the data they had was sufficient to be worth pursuing their abstraction goal at all (C1).

% 3. Clinical concepts of interest vary in consistency and level of thoroughness in documentation; participants want ways to quantify and be alerted to this uncertainty.

\subsection{Specifying Tasks: Managing Complexity in Nuanced Abstraction Goals}
\label{sec:specifying-tasks}

For many of the tasks we attempted to define, the researchers in our study had already performed some initial manual chart review (B, E, F) or even completed labeling a small dataset (A, C, D, G) for the task.
We therefore assumed that specifying these tasks in Libretto would be a simple matter of articulating to the LLM the instructions they had followed in their own chart review process.
Instead, we found that even after multiple sessions working with Libretto over the course of months, it remained difficult to precisely capture the researchers' task definitions due to the following challenges:

% We were ultimately able to define at least one usable abstraction specification for each project except project F (which was not able to participate in Phase 2 of the study due to technical infrastructure delays).
% Despite this overall success, in most cases clinicians were not able to complete the specification process without data scientist guidance, due to the challenges described below.

\subsubsection{Even simple tasks require clinical nuance to implement.}

Through the process of specifying abstraction tasks, we were surprised to learn that the perceived ease of abstraction of a concept had little correlation with the difficulty of obtaining it with an LLM.
As described in Section \ref{sec:reviewing-data}, participants came in with strong intuitions about tasks that would be easier for the LLM to extract based on their knowledge of what was typically documented in notes (A1, B1, C2, D1).
For example, researcher D1 initially expected that extracting the stage of a patient's cancer for Task 9 would be relatively simple: \textit{``I mean, like, }I\textit{ know where to look.''}
They considered clinical histories and decision-making rationales to be among the hardest for an automatic tool to label since they would require inference and judgment.

In reality, however, the latter abstraction tasks ended up being the simplest to extract and required very little refinement to the specification because the LLM could simply output free-form text.
Participants using these types of prompts were impressed by the LLM's ability to infer useful information, such as G1 who observed the LLM inferring a patient's residence in temporary housing as a barrier to scheduling a colonoscopy for Task 14: \textit{``Oh! So they found his listed residence. That's good.''}
Notably, we did not observe any occurrences of hallucination across any of the seven projects, indicating that the LLM was accurate enough to output answers that were true based on the notes (even though it might not always obey the format requirements).

On the other hand, cancer stage proved to be one of the most challenging concepts to extract because of the nuances in defining cancer stage to begin with.
All three tasks attempting to extract staging information (Tasks 6, 9, and 11) faced challenges around (a) different descriptions by different types of providers (e.g., pathological stage versus clinical stage), and (b) varying stages over time for the same patient.
For example, E1 wanted to extract \textit{``the stage at cancer diagnosis, and also at the time of the most recent note and visit, because sometimes they're diagnosed with an early stage and then later get diagnosed with metastatic disease.''} 
We were unable to capture this timing information directly with the LLM and instead had to manually post-process the extractions in code.

Another fundamental limitation that arose with the existing specification structure was that many important concepts required synthesizing information across multiple notes.
Many prior information extraction approaches have focused on capturing values one note at a time~\cite{aronson_overview_2010,balasubramanian_leveraging_2025,williams_using_2026}.
Indeed, LangExtract's approach was to simply scan notes chunked by length and return extractions independently, assuming that each extraction would hold true across the entire patient timeline.
% This worked well for extracting information in pathology reports and decision-making rationales, which were often directly stated in a single note; indeed, previous literature has explored these types of extraction extensively~\cite{balasubramanian_leveraging_2025,williams_using_2026}.
However, this method fell short for several specifications in our study, reflecting the more demanding requirements of conducting this work within real research projects.
In particular, participants working on cancer recurrence or progression described needing to triangulate or compare multiple sources to verify timing or differentiate \textit{suspicion} and \textit{confirmation} of the outcome.
Collaborator F1 described that recurrence \textit{``may come from, say, a biopsy which showed it, or a scan which showed it, or a clinician who documented it. Then you would have to actually use different data sources at the same time to save... a first date of recurrence.''}
Similarly, progression and changes in tumor characteristics would require the LLM to qualitatively and quantitatively compare extractions at multiple time points,  which clinicians agreed would be highly useful but would require a complex interaction between prompts and code (A1, E1).

\subsubsection{Hard to achieve transparency and control in prompt writing, either manually or using vibe-coding.}

The LangExtract specification format allowed users to type free-text prompts, schemas, and examples, ostensibly providing them full control over the output and ensuring LLM reliability.
However, several participants found the schema and examples confusing (D1, E1, G1) and excessive (F1), and avoided editing them altogether:
\begin{quote}
    \textit{``If I wanted to edit [the prompt], it would be intuitive, but are [we] supposed to also be looking at the schema, and editing the schema, and also the examples...? I don't know where this is coming from. I started eyeballing it just to make sure it looked, like, reasonable.''} (E1)
\end{quote}

The ``Generate Spec'' feature was perceived by some to solve this problem because it could easily automate the task of filling out the complex specification fields.
Some participants were able to obtain a prompt that worked well on the first try by simply requesting the data elements they wanted. 
For Task 5, E1 typed, \textit{``I want to know MMR and HER2 status from path reports''} (referring to genetic markers that can indicate how tumors will respond to treatment), and received a prompt that was immediately usable.
On the other hand, the same participants sometimes found that the generated specs created unforeseen problems when run on real notes, most often around generating excessive or spurious outputs (C1, C2, D1, E1).
For example, C2's initial attempt at a prompt to extract tumor information from pathology reports produced over 300 extractions for a single patient, most of which were redundant.
In other cases, because LangExtract divided the notes into chunks to process them within the LLM's context window, it produced extractions indicating negative results when there was no information relevant to the task in a given chunk.
These implementation-specific issues delayed the researchers' ability to focus on more clinically relevant questions.
Participants also found it difficult to edit the specifications themselves because the generated prompts were also redundant, including the same instructions across the prompt, schema, and examples.
As E1 noted: \textit{``Initially... I would just edit the prompt directly, but then I realized that... whatever I wrote here didn't carry over into the schema or the examples,''} which led them to use the Generate feature exclusively to update all fields at once.
As a result, the researchers ended up relying on a generative tool into which they had limited visibility or control, contrary to the goal of giving them \textit{more} control over the abstraction definition.

Given this feedback, during the second phase of the design process, we prototyped an alternative specification style: a structured form where clinicians could directly articulate structured ``variables'' they wanted to extract, with free-form instructions only used to supplement the structured definitions.
Participants were interested in this format because it could scaffold their process of writing the specification themselves: \textit{``it's helpful to know, these are the things to include in your schema or ... your prompt''} (D1).
However, they also expected that they would want to vibe-code even these more straightforward specs because the LLM could do the heavy lifting of formalizing the abstraction task; as E1 put it, \textit{``I like that it decides for me.''}
These findings suggest that although both prototypes were designed to provide both AI automation and user control over specifications, neither one offered the right balance in practice.

% IDEA: have a table with all the extraction goals pursued and the issues we identified with each

% 1. Participants have intuitive ideas to operationalize and formalize difficult extraction goals, but the complexity of these tasks when translated to LLM instructions is difficult to manage.

% 2. Participants find the declarative LangExtract spec format confusing and instead prefer to “vibe code” specs. However, the resulting specs tend to require iteration because they produce overwhelming and redundant results, and they are hard to edit.

\subsection{Evaluating Results: Balancing Ad-Hoc Clinical Judgment with Analytical Consistency}

This final stage of the Libretto workflow was where clinicians brought the most experience because of their exposure to pre-LLM chart review.
While they saw potential for the LLM to scale up their data abstraction goals, they also expressed concerns about how they would know if the tool was sufficiently accurate for their needs, given the lack of already-available ground truth labels.

\subsubsection{Reliance on manual effort and clinical expertise.}
Most collaborators in the study had experience manually reviewing patient charts to abstract information, either for the current project or a previous one (A1, C1, C2, D1, E1, G1, G2).
In fact, chart abstraction was often seen as a core component of medical research training, where medical students or research coordinators would learn the ins and outs of patient notes while contributing to research projects (B1, F1).
At times this process requires substantial clinical judgment and expertise; as F1 described:
\begin{quote}
    \textit{``A common [challenge] is performance status, how well the patient is. It's just not there anywhere, and then you have to read through the notes and the narratives to kind of figure out... If I have a student with me, I'll train them - how do you collect performance status, right? I'll say, `okay, look at the most recent clinical notes, look to see if they're using any walking aids, look to see if they're in a nursing home.' So, I give them that sense of how to collect it, even though if it's not written down.''} 
\end{quote}

Participants were skeptical that this trained process of deliberation could be extended to an LLM, and often expected that human involvement would be necessary to correct errors and discrepancies.
Clinicians described heuristics that they might use to decide what output to trust if there were two different values, including the type of note (A1, C2, D1), the date it was documented (D1), or even whether the result was found at the researcher's institution versus from an external provider (D1, E1).
However, these patterns were viewed as case-by-case and ultimately up to the researcher: \textit{``I'm not sure how you would have an automated way that actually does the adjudication, right? Because a human would always have to double check''} (C1).
They also noted that different clinicians and projects might decide on different policies for resolving conflicting information.
Libretto's chart abstraction specifications would be a promising place to document such policies to improve the reproducibility of research, but they could also become overly complex and unwieldy due to the challenges described in Section \ref{sec:specifying-tasks}.

% 1. Ad-hoc clinical judgment on chart review takes significant time but is a core part of the process that often requires engagement from both project leaders and trainees doing labeling. Clinicians see potential in using the LLM to scale up this niche set of skills.

% 2. Resolving multiple conflicting answers for the same extraction question often requires context-dependent judgment and clinical expertise, which can be difficult to articulate to an LLM.

\subsubsection{Need for pragmatic, ad-hoc evaluation on specific patient subtypes.}
Participants found the process of evaluating LLM outputs in the Extractions view intuitive, and they particularly appreciated the ability to jump to the context of an extraction: \textit{``that makes it easy to verify''} (F1).
However, two main challenges arose with Libretto's current evaluation workflow.
First, clinicians often described specific criteria for patient characteristics that would make effective tests for the model, such as patients with particular types of specialist notes (A1) or those with changing cancer stages: \textit{``I'd love to see, like, what it shows for someone who had... initial stage 2, and then later developed stage 4''} (E1).
This type of semantic search would be difficult to implement without sophisticated text understanding, yet scaling a language model to search across thousands of patients and notes at interactive speeds is non-trivial. 

The second, more fundamental challenge was rigorously estimating the quality of a model's output for a specification and deciding whether it was ``good enough'' to run across all patients.
The gold-standard approach for validating an automatic chart abstraction task---manually labeling a subset of patients and quantifying model accuracy---is a major time and resource bottleneck.
Participants found that the LLM could essentially produce human-quality annotations as long as the specification was sufficiently good quality (A1, E1), so they instead preferred ad-hoc methods that allowed them to quickly identify ways to improve the spec.
F1 described their approach to validating prompts: \textit{``I basically will do 5 at a time... and I'll verify, so once my prompt is at a point where I feel like I got exactly what I want, then I would just run 100.''}
However, there was no way to quantify accuracy under this adaptive evaluation setup (A1, F1), making it difficult to know when to switch between evaluating results and re-specifying the task (A1, E1).
As a result, despite participants' perceptions that the outputs were high-quality, more formal, resource-intensive evaluation would currently be required to ensure these specifications were robust enough for publication-quality results.
Furthermore, it was unclear whether these specifications would generalize across institutions or if they would be limited by context-specific data characteristics.

% 3. Validating specifications at the level of a few patients is a crucial and difficult step. Clinicians want the system to help them efficiently find and review patients with specific characteristics efficiently, audit the LLM output on these patients, and cross-reference them with who and what they see as clinicians.

\section{Discussion}

This work presented a longitudinal collaborative effort to understand the barriers to making LLM-assisted data abstraction workflows feasible for clinical research.
Clinical data abstraction is an underexplored task in HCI research, and our work addressed this gap by bridging existing literature in AI for healthcare~\cite{feng_human-ai_2026,li_endoextract_2026,agrawal_large_2022} and AI-assisted data analysis~\cite{kazemitabaar_improving_2024}.
We adopted a longitudinal co-design approach across multiple clinical contexts, allowing us to identify practical, real-world patterns of clinician-LLM interaction.
% Compared to classical methods for clinical IE, LLMs create fundamentally different patterns of use with different opportunities and associated challenges.
% The Libretto workflow allowed the clinical researchers in our study to explore nuanced and subjectively-defined concepts, such as cancer stages or rationales for provider decisions, unlike classical IE methods which are largely limited to predefined concepts and explicit text mentions.
Across Libretto's workflow of defining abstraction tasks---reviewing data, specifying a task, and evaluating the results---clinicians faced difficulties formalizing their desired concepts, which were often more complex than those previously explored in medical informatics research~\cite{choi_developing_2023,balasubramanian_leveraging_2025}.
We also found that clinician researchers needed lightweight ways to evaluate LLM-based abstraction to ensure a correct, robust specification without time-consuming manual labeling. 
We first contextualize these findings against prior literature in human-AI interaction, then suggest a range of future design directions to address the identified barriers.

\subsection{Challenges in Human-AI Collaborative Clinical Research}

A primary implication of this work is that for many clinically-meaningful abstraction tasks, the specification process is much broader and more difficult than writing a prompt.
The clinical IE literature predominantly considers extraction tasks as fixed and ignores considerations such as note and provider reliability or triangulation across multiple notes for simplicity~\cite{hsu_llm-ie_2025,choi_developing_2023,williams_using_2026,agrawal_large_2022}.
Moreover, the conventional ``iterative'' approach has been to allow clinicians to modify the prompt(s) while holding other factors constant~\cite{feng_bayesian_2025,hein_iterative_2025}.
By directly studying how clinical research teams define abstraction tasks from the ground up for real-world research goals, we saw that \textbf{iterative refinement is often necessary across the \textit{entire} data pipeline, from patient cohort selection to note filtering to prompting and synthesis across extracted results.}
Moreover, these refinements draw extensively on clinical intuition and judgment, so they often cannot be anticipated or performed by members of the team without domain expertise.

Our work expands on a line of recent work in HCI that aims to make AI-assisted data analysis practical and accurate for people without programming expertise.
It is well-understood that despite the strong potential of LLMs to assist in or automate data science work~\cite{hong_data_2025,sun_lambda_2026}, it can also be challenging for users to verify, critique, and control their outputs~\cite{obrien_how_2025,drosos_its_2024}.
The ``tools for thought'' literature addresses this challenge with shared task abstractions that both users and LLM agents can interpret and control, by translating code into a graphical or plain-language representation~\cite{kazemitabaar_improving_2024,ma_tempoql_2025}.
\textbf{While it is often assumed that a natural-language prompt to be run over a set of documents is an inherently interpretable artifact~\cite{feng_human-ai_2026,lam_concept_2024,shankar_docetl_2025,goel_introducing_2025}, our study showed that this is not always the case.}
Clinicians neither authored specifications directly in Libretto nor made major edits after vibe-coding an initial draft, even when the generated spec was misaligned with their goals; it was often challenging to communicate the desired changes without first discussing with the data science team.
This may have occurred because the specification structure itself (prompt, schema, and examples, as used by LangExtract~\cite{goel_introducing_2025}) was not aligned with clinicians' mental models of how clinical data abstraction should work.
Our prototype expressing the specification as a list of variable definitions improved alignment with participants' conceptualizations but sacrificed flexibility and ease of entry, suggesting that a hybrid of these approaches might be needed.
Improving the transparency of specifications, including both the prompts and adjacent data choices, will become especially important as complex abstraction tasks (e.g., cancer recurrence) become more common in clinical research.

This work also speaks to the ongoing challenge across machine learning research of how to efficiently and rigorously validate task-specific AI outputs.
While it was often straightforward to define evaluations in classical ML research because the methods already required labeled data, LLM-based methods can achieve high accuracy with few or no labeled examples~\cite{agrawal_large_2022,wornow_2025_zero}.
Even so, labeled data has remained necessary to evaluate these methods' accuracy, since LLMs can struggle with edge cases that appear in held-out datasets despite performing well in spot-checks~\cite{liu_human_2025}.
\textbf{Our study suggests that clinicians could effectively red-team healthcare AI models~\cite{pan_addressing_2026} by actively searching for failure modes, but they currently lack tooling and structured frameworks to do so efficiently and rigorously.}
Participants described specific types of patients on which they suspected their prompts might fail, yet Libretto did not support finding them because the search criteria were themselves complex and unstructured.
There was also a steep time cost associated with reviewing patients due to the volume of notes for each case; this cost would be even higher when reviewing a full evaluation dataset.
Finally, the adaptivity of this red-teaming process makes it difficult to quantify the accuracy of a model on a spec, which is crucial to establish the validity of the data when it is used in a downstream clinical research study.
% Active learning~\cite{kholghi_active_2017} and testing~\cite{ribeiro_adaptive_2022,zhou_adaptive_2026} may address this issue by enabling early stopping under a predefined sampling scheme, but it remains unclear how experts can steer the sampling process while still obtaining statistically sound accuracy estimates.
% These findings are especially important as specification-driven development, where the specification is the primary human-authored artifact rather than the code, rapidly becomes the predominant workflow in software development and data science.
% Tools such as GitHub's Spec Kit~\cite{delimarsky_spec-driven_2025} and Plan mode in Claude Code\footnote{\url{https://code.claude.com/docs/en/permission-modes}} enable this workflow 

% \begin{enumerate}
%     \item AI-assisted data analysis and sense-making; tools for thought, intermediate artifacts
%     \item Specification-driven development and agentic prompt optimization
%     \item Evaluation and active learning
% \end{enumerate}
% What about agentic prompt optimization? Why wouldn't that work?

\subsection{Future Design Directions}

Our work reveals several areas where further research is needed to develop and design human-AI collaborative workflows, both for clinical data abstraction and the broader needs of domain expert researchers. 
We describe these directions below according to the stages of the Libretto workflow where they arose:

\subsubsection{Reviewing Data}
\begin{enumerate}[label=\textbf{D\arabic*.},leftmargin=*]
    \item \textbf{Agentic cohort selection.} The Reviewing Data stage in Libretto's current design required significant and time-consuming iteration between clinical collaborators and the data infrastructure experts on the team, to ensure that the abstraction process was being applied to the most relevant set of patients and notes. This process has the potential to be automated by AI coding agents~\cite{ma_tempoql_2025}.
    However, it is unknown whether these tools could fully substitute for human expert data scientists in healthcare databases, which are notoriously complex and heterogeneous.
    \item \textbf{Incorporating other data modalities.} Beyond text notes, participants in our study often alluded to other forms of data that could provide useful signal to their abstraction tasks, such as medication lists, socioeconomic attributes, and radiological images. Careful integration of these data sources into the patient review process and as inputs to the LLM could help overcome inconsistencies in text-based documentation.
    \item \textbf{Presentation of patient trajectories.} While seemingly simple, visually representing the details of a patient's healthcare journey has been a long-standing challenge across HCI and medical informatics~\cite{Ghassemi2018,sultanum_more_2018}. While LLMs can often accurately summarize clinical notes~\cite{van_veen_adapted_2024}, integrating information from across hundreds of documents while maintaining contextual relevance remains challenging~\cite{kothari_when_2026} and would provide a significant speed boost for clinicians reviewing and evaluating patients.
    \setcounter{designCounter}{\value{enumi}}
\end{enumerate}

\subsubsection{Specifying Tasks}
\begin{enumerate}[label=\textbf{D\arabic*.},leftmargin=*]
\setcounter{enumi}{\value{designCounter}}
    \item \textbf{Optimizing specification structure for transparency and control.} As described in Section \ref{sec:specifying-tasks}, we found that neither the free-form LangExtract specification structure nor a more structured variable definition list were sufficiently intuitive for researchers to easily use, leading them to vibe-code specs with minimal edits. An important direction for future research is to determine what specification designs \textit{do} support domain expert control, such as graphical representations (e.g. decision tree diagrams), example inputs and outputs, or spreadsheet metaphors. Our results suggest that the optimal spec structure should be \textit{aligned} with researchers' experiences with manual chart review, \textit{non-redundant} to facilitate editing, and \textit{high-level} while capturing clinically-relevant nuances.
    \item \textbf{Orchestrating the synthesis of extraction results.} In many of the abstraction tasks our collaborators explored, the desired final result was not a direct extraction from a note but rather a custom combination of multiple pieces of evidence (e.g., combining pathologic and clinical evidence to confirm cancer recurrence). While some document processing tools support the creation of such multi-step workflows~\cite{shankar_docetl_2025}, they require technical expertise to implement and understand the underlying execution plans. Bringing these sophisticated pipelines to non-technical domain experts would likely require novel interaction designs.
\setcounter{designCounter}{\value{enumi}}
\end{enumerate}

\subsubsection{Evaluating Results}
\begin{enumerate}[label=\textbf{D\arabic*.},leftmargin=*]
\setcounter{enumi}{\value{designCounter}}
    \item \textbf{Efficient patient search and matching.} One direction for future research to which LLMs are particularly amenable is in helping clinicians find relevant patients to review. Patient record matching has been explored extensively in the context of matching patients to clinical trials~\cite{wornow_2025_zero}, yet other types of patient similarity have been less explored in medical informatics~\cite{sharafoddini_patient_2017}. Given that quickly identifying an example of a failure mode or edge case could greatly accelerate researchers' specification process, LLM-based embedding and similarity search over patient trajectories could be a valuable addition to the clinical data abstraction workflow. Compared to existing retrieval augmented generation (RAG) approaches, it would be important to ensure that this automatic search process captures temporal aspects of patient records (e.g., diagnosis, then treatment, then progression) and maintains efficiency in a large patient cohort.
    \item \textbf{Communicating uncertainty and conflicts.} Participants expressed a range of context-specific methods for resolving data availability and reliability concerns, including prioritizing specific note or provider types or indexing on certain landmark time-points (Section \ref{sec:reviewing-data}); however, they lacked the ability to systematically investigate the prevalence of these issues. Building on prior research in uncertainty quantification~\cite{turner_linguistic_2021}, future data abstraction tools could automatically surface potential inconsistencies, conflicts, and sources of uncertainty that users could use to refine their specification.
    \item \textbf{Active, user-guided validation.} Ad-hoc evaluation is increasingly common in AI work~\cite{perez_red_2022} and was preferred by the researchers in our study. However, rigorous accuracy estimation and identification of failure modes continue to be paramount for assessing the validity of clinical research~\cite{reichenpfader_consensus-based_2026}. Some recent work has explored how frameworks such as active learning~\cite{ribeiro_adaptive_2022,abbas_clinical_2024,kholghi_active_2017} and safe anytime-valid inference~\cite{zhou_adaptive_2026} can be used to structure these adaptive workflows; however, future work is needed to understand how these techniques can be applied to highly complex patient notes and abstraction tasks. %Future work is also needed to design human-AI collaborative workflows to identify test cases and implement specification improvements while measuring model performance.
\end{enumerate}

\subsection{Limitations}

One limitation of this work was that our design artifact focused on eliciting clinician researchers' \textit{perceptions} of model output quality, rather than quantitatively evaluating quality through a blinded annotation process.
Given the high time cost of manual labeling and the time spent resolving specification issues during the study, this type of validation was left as an important area for future work to evaluate Libretto.
A related limitation is that we kept the Libretto design mostly consistent with prior systems for clinical IE (such as LangExtract) throughout the co-design process~\cite{burdenko_medical_2024,goel_introducing_2025}. 
This choice allowed us to evaluate gaps in the workflows supported by currently-available tooling, while participants' suggestions for more sophisticated LLM-driven features serve as directions for future work.

This work focused on particular medical and institutional contexts, albeit with the involvement of multiple research teams.
While cancer research poses distinct challenges for clinical data abstraction, such as highly longitudinal patient records, heterogeneous patient characteristics, and multi-disciplinary care, these issues likely pertain to many other conditions as well.
We also conducted this research within two academic institutions where researchers are generally positive on data sharing and LLM use in medicine.
It is important to explore how the needs of clinician researchers might differ in other settings, such as institutions where AI adoption has been more limited or where medical documentation practices are more variable.
    % 1. Did not quantitatively evaluate the performance of the prompts that participants developed due to the resource cost - this is an important area of future work
    % 2. Participants were at institutions that were predominantly positive on LLM use and had ready access to LLM APIs
% So far, LLMs have been widely tested for various clinical data abstraction tasks, but the 

\section{Conclusion}

Data abstraction from patient records has been an essential tool in clinical research over the past several decades, shedding light on patterns from individual patient experiences that can reveal the underlying causes of adverse outcomes and point to promising treatment practices.
LLMs have opened opportunities to glean even more rich insight across ever-growing healthcare databases, and our work is among the first in HCI to explore where these tools succeed and fail in supporting the abstraction process.
Our findings highlight that despite the promising capabilities of LLMs in extracting clinical information, there remain significant challenges in specifying abstraction tasks to yield outputs that are meaningful, accurate, and aligned with expert clinical judgment.
Through technical innovations in AI systems and novel human-centered designs, we envision LLM tools being applied in responsible and robust ways to help clinical researchers advance our shared knowledge about human health.

\begin{acks}
The authors thank the many clinical researchers, data analysts, and research assistants who either directly used Libretto or helped us define abstraction tasks and process the patient data. This work was supported by the Weill Cancer Hub West and the Weill Family Foundation.
\end{acks}

\bibliographystyle{ACM-Reference-Format}
\bibliography{main}

% \appendix

\end{document}